# Probe to properties of MgB$_2$ thick film on silicon carbide substrate

Fen Li, Tao Guo, Kaicheng Zhang, Li-ping Chen, Chinping Chen and Qing-rong Feng*

Department of Physics and State Key Laboratory for Artificial Structure and Mesoscopic Physics

Peking University, Beijing 100871, P.R. China

## Abstract

We have successfully synthesized MgB$_2$ thick films on 4H-SiC substrate by hybrid physical-chemical deposition (HPCVD). The films have transition temperature $T_C$ above 40 K. X-ray diffraction (XRD) shows the $c$-axis oriented structure of MgB$_2$, with Mg and small MgO impurities. The critical current density $J_C$, estimated using the measured magnetic hysteresis loop and the Bean model, is 6 MA/cm$^2$ in self field at 10 K.

The superconductivity in MgB$_2$ has generated great interest in its applications in superconducting electronics [1] and high magnetic fields [2]. It is hoped that reproducible and uniform Josephson junctions may be easier to fabricate using MgB$_2$ than high temperature superconductors. The transition temperature of 39 K allows operation of MgB$_2$–based circuits at above 20 K, much higher than the liquid helium temperature. Hence, the required cryogenic environment can be conveniently obtained using crycoolers without the cumbersome facilities necessary to handle the cryogenic liquid. The benefits are obvious, immediate and very attractive for superconducting integrated circuits. For high magnetic field superconductors, various techniques have been applied, including powder-in-tube (PIT) [3-6], diffusion of Mg vapor into B-fiber [7], etc, in an effort to find out an effective and efficient process to produce superconducting wires and tapes. Very impressive successes have been achieved. However, alternative approaches are also under intensive studies in the laboratories. One of the routes may rely on the fabrication of thick films on various substrates. For example, thick MgB$_2$ films deposited on graphite [8-9], stainless steel [10], etc., by HPCVD have been reported with the film thickness ranging from 1 to over 40 μm. The Penn. State groups have successfully grown MgB$_2$ epitaxial thin film, less than a few hundred nanometers, on SiC substrate [11]. High–quality MgB$_2$ thick films grown on SiC substrate by an in-situ deposition process are important for a better understanding of their basic properties. In this report, we present the results of MgB$_2$ thick films grown on SiC substrate by HPCVD and their excellent properties.

Recently, we have succeeded in depositing MgB$_2$ thick film with an average thickness of about 2.5 μm on the 4H-SiC substrate. It is based on the techniques of HPCVD using the mixture of 75% B$_2$H$_6$ in H$_2$ and Mg ingots as the active sources, similar to that reported previously [12]. Additional pure gas of H$_2$ also flows in the reaction chamber, serving to reduce the oxygen content and to suppress any further oxidation of the sample during the deposition process. Also, the H$_2$ would cut down the decomposition rate of B$_2$H$_6$. The flow rate of the B$_2$H$_6$ mixture gas was about 10 sccm at the pressure of 2 KPa and the background gas mixture, $H_2$+Ar (90%), was about 100 sccm at 18 KPa. The temperature of the chamber was controlled within the range of 697~703 °C. Under these conditions, the film deposition rate is about 2.1 nm/s.

The thick $MgB_2$ film was characterized by scanning electron microscopy (SEM) using a QUANTA 200 FEG scanning electron microscope. The general feature of the film surface is shown in figure1. Figure 1(b) is a magnified view of figure 1(a). From figure 1(a) we learn that though the film surface is not very smooth, it is quite dense. Figure 1(b) shows that the film is composed of regular hexagonal $MgB_2$ crystallites, most of which are $c$-axis oriented. This can also be seen in the XRD pattern shown in figure 2. The average size of the crystallites is about 300 nm. Figure 1(c) shows the cross sectional image of the film. The two arrows mark the thickness of the film. The average thickness is about 2.5 μm.

The XRD was performed using a Philip X'pert diffractometer with a $K_\alpha$ radiation source. The spectrum, presented in figure 2 using a logarithmic scale on Y-axis, shows a mostly $c$-axis oriented polycrystalline $MgB_2$ crystal structure with the presence of some impurities, such as MgO and Mg. Except for the (001), (002) main peaks of $MgB_2$, there is a very small (201) peak in the spectrum. It suggests that the film is with mostly $c$-axis oriented structure. The Mg peak is higher than that of the MgO, indicating that there is more Mg than MgO in the film. Due to the excessive Mg vapor during the depositing process, Mg can be present in the deposited films. Furthermore, the presence of MgO may be resulting from the residual $O_2$ adhering to the chamber inner-wall and the surface of Mg ingots, existing in the background mixture gases, $H_2$+Ar, etc.

The temperature-dependent resistance (*R-T*) curve was performed by the standard 4-probe measurement using a Quantum Design PPMS System. The transition temperatures in self field are determined as $Tc$(onset) = 40.5 K and $Tc$(zero) = 39.5 K. The upper critical field $H_{C2}(T)$ curve, figure 3, is obtained by a series of $Tc$(onset) from the *R-T* measurements under applied magnetic fields up to 8 T which are shown in the inset of figure 3. The data points thus determined can be best fit by the polynomial function, $H_{C2}(T) = a_0 + a_1T + a_2T^2$. The extrapolation to $T = 0$ K reaches as high as 40.8 T. In comparison with the $H_{C2}(T)$ of the clean $MgB_2$ film ~ 1.3 μm on graphite[9], the present film has a much higher value at the corresponding temperature. For example, within the range of experimental measurement, $H_{C2}$ reaches 8 T at $T = 25$ K for the present film and at $T = 17$ K for the clean film on graphite [9]. By extrapolation to $T = 0$ K, $H_{C2} \sim 13.7$ T for the latter, much smaller than the present one. Though we can not confirm the amount, distribution and particle size of MgO, the work of P. Kováč *et al* has shown that the well distributed and small sized MgO particles improve flux pinning [13].We believe that the presence of MgO and Mg in the deposited film can be self optimized between the inter-grain connectivity and the pinning effect. This would in turn improve the upper critical field $H_{C2}$.

At the same time, a series of *M-H* measurements were carried out with the applied magnetic field $H$ parallel to the film surface using a Quantum Design SQUID magnetometer. The results are plotted in the inset of figure 4(a). At $T = 5$ K under low applied magnetic field, especially in the 0.05 ~ 0.24 T region indicated by the arrow, a phenomena of flux jump occurs. This is related to the very fine disorder structure and the relatively small thermal diffusion in the films at low temperature [14]. The derived critical current density $Jc$ is shown in figure 4 as a function of temperature under $H = 0$, 0.5, and 1.0 T. It is calculated according to the Bean Model, $Jc = 30 \Delta M/r$ [15], where $\Delta M$ is the width of the magnetic hysteresis loops, $r$ is the equivalent radius determined by $\pi r^2 = ab$. $a = 4.7$ mm, $b = 4.7$ mm are the length and width, respectively. In the zero field $Jc$ is 6 MA/cm$^2$ at 10 K. In the flux jumping region at $T$ = 5 K, the Bean Model can not apply. So we obtain the corresponding $Jc$ ($T$) value by extrapolation from a best polynomial fitting as 7.4 MA/cm$^2$. These are values comparable to the reported $Jc$ [6,11,16,17]. In the low applied field at $H$ = 0.5 T, $J_C$ is suppressed to 2.4 MA//cm$^2$. We speculate that the quick suppression of $Jc$ by the applied magnetic field is likely attributed to the small amount of MgO pinning centers in the film. Further study in this respect is under way.

In conclusion, the dense MgB$_2$ thick films with an average thickness of 2.5 μm have been effectively grown by the technique of HPCVD on 4H-SiC substrate. The transition temperature $T_C$ (onset) = 40.5 K with $\Delta T$ = 1 K. X-ray diffraction shows a highly $c$ - axis oriented structure. The film contains impurities of MgO and Mg. These may act as flux pinning centers, resulting in a high upper critical field at 0 K, $H_{C2}$(0) = 40.8 T, which is determined by an extrapolation from a best polynomial fit to the data. By the magnetic hysteresis loops and using the Bean model, we get the critical current density, $J_C$ = 6 MA/cm$^2$, in self field at 10 K and 7.4 MA/cm$^2$ at 5 K.

## Acknowledgement

We thank Dr. Xiaoxing Xi for helpful discussions and his comments on the manuscript. This work is supported by NSFC under contract No. 50572001. We appreciated the technical supports of CHEN Li, and WANG Yong-zhong.

Figures

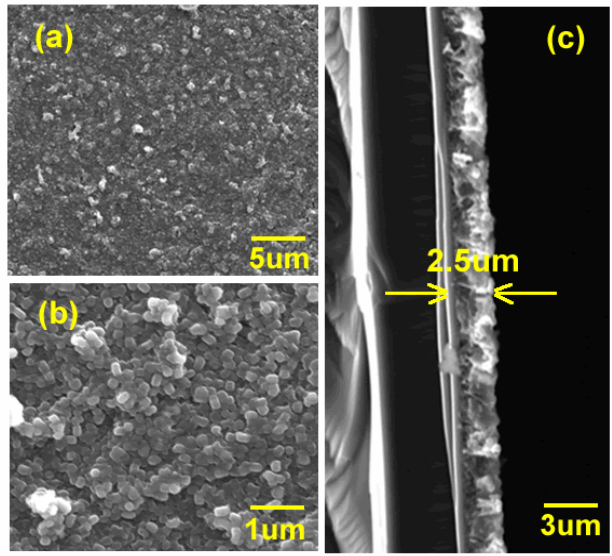

Figure 1. SEM images of the film surface. (a), (b) are the morphologies with the magnification of 3000 and 30000, respectively. (c) is the cross-section of the grown MgB$_2$ film.

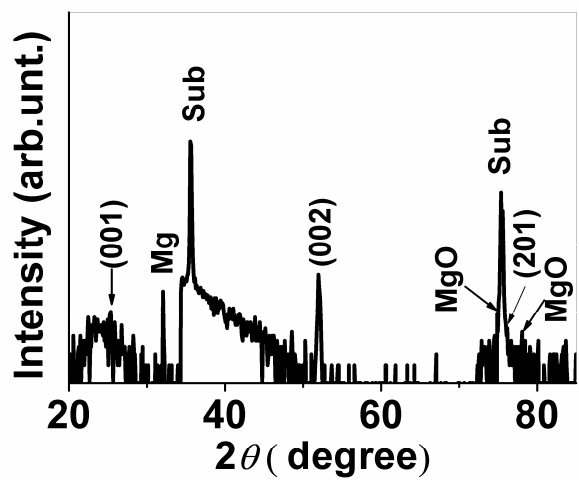

Figure 2. XRD spectrum of the MgB$_2$ thick film on 4H-SiC substrate.

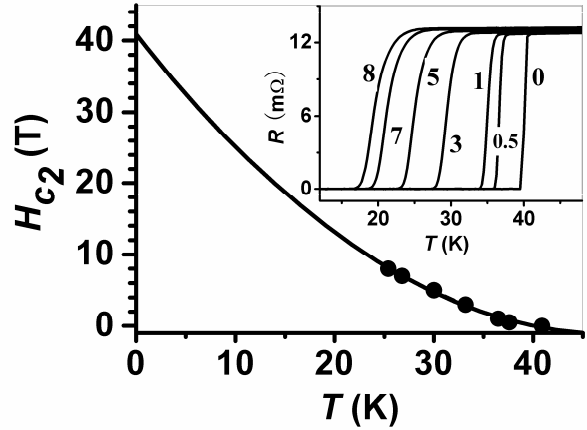

Figure 3. Upper critical field $H_{C2}$ as a function of temperature. The solid circles are data points determined from the *R-T* measurements shown in the inset. The solid curve is the fitting line for the polynomial function, $H_{C2}(T) = a_0 + a_1T + a_2T^2$. The applied fields for the *R-T* measurements range from 0 up to 8 T, indicated by the number in the inset for each curve.

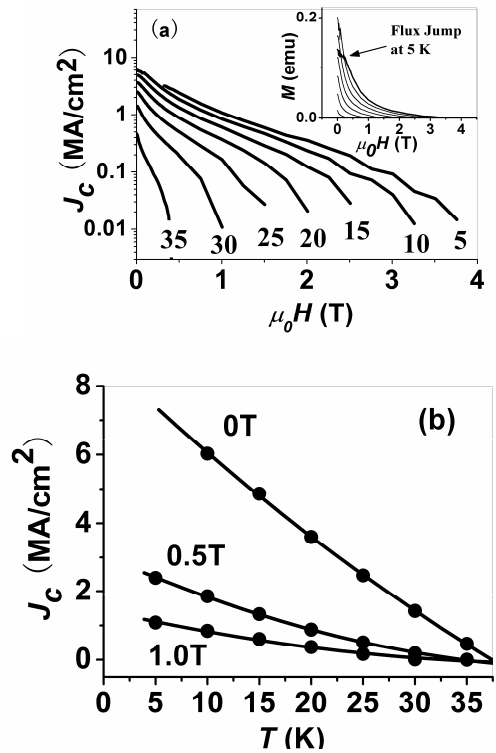

Figure 4. (a) Critical current density *Jc* versus the applied field at *T* = 5, 10, 15, 20, 25, 30, 35 K, as marked by the numbers. *Jc* is determined by the calculation according to the Bean model on a series of *M-H* measurements shown in the inset. (b) *Jc* as a function of temperature in different applied magnetic fields